\documentclass[sigconf, screen]{acmart}
\AtBeginDocument{%
  \providecommand\BibTeX{{%
    \normalfont B\kern-0.5em{\scshape i\kern-0.25em b}\kern-0.8em\TeX}}}

\setcopyright{rightsretained}
\acmPrice{}
\acmDOI{10.1145/3533767.3534367}
\acmYear{2022}
\copyrightyear{2022}
\acmSubmissionID{issta22main-p20-p}
\acmISBN{978-1-4503-9379-9/22/07}
\acmConference[ISSTA '22]{Proceedings of the 31st ACM SIGSOFT International Symposium on Software Testing and Analysis}{July 18--22, 2022}{Virtual, South Korea}
\acmBooktitle{Proceedings of the 31st ACM SIGSOFT International Symposium on Software Testing and Analysis (ISSTA '22), July 18--22, 2022, Virtual, South Korea}

\usepackage{verbatim}
\usepackage{amsmath}

\usepackage{amssymb}
\usepackage{algorithmic}
\usepackage{algorithm}
\usepackage{makecell}

\usepackage{graphicx}
\usepackage{epstopdf}
\usepackage{array}
\usepackage{booktabs}
\usepackage{multirow}
\usepackage{caption} 
\usepackage{multicol}
\usepackage{tabularx}
\usepackage{framed}
\usepackage{enumitem}
\usepackage{threeparttable}

\graphicspath{{figs/}}

\usepackage{xcolor}

\usepackage{colortbl} 
\definecolor{jade}{rgb}{0.0, 0.66, 0.42}
\definecolor{carolinablue}{rgb}{0.6, 0.73, 0.89}
\definecolor{dkgreen}{rgb}{0,0.6,0}
\definecolor{dkblue}{rgb}{0,0.4,0.5}
\definecolor{gray}{rgb}{0.5,0.5,0.5}
\definecolor{mauve}{rgb}{0.58,0,0.82}

\definecolor{codegreen}{rgb}{0,0.6,0}
\definecolor{codegray}{rgb}{0.5,0.5,0.5}
\definecolor{codepurple}{rgb}{0.58,0,0.82}
\definecolor{codeblue}{rgb}{0,0,205}
\definecolor{backcolour}{rgb}{245,245,245}

\usepackage{listings}
\usepackage{enumitem}

\lstdefinelanguage
   [x64]{Assembler}     
   [x86masm]{Assembler} 
   {morekeywords={xend, CDQE,CQO,CMPSQ,CMPXCHG16B,JRCXZ,LODSQ,MOVSXD, %
                  POPFQ,PUSHFQ,SCASQ,STOSQ,IRETQ,RDTSCP,SWAPGS, %
                  rax,rdx,rcx,rbx,rsi,rdi,rsp,rbp, %
                  r8,r8d,r8w,r8b,r9,r9d,r9w,r9b, %
                  r10,r10d,r10w,r10b,r11,r11d,r11w,r11b, %
                  r12,r12d,r12w,r12b,r13,r13d,r13w,r13b, %
                  r14,r14d,r14w,r14b,r15,r15d,r15w,r15b}} 

\lstdefinestyle{mystyle}{
    backgroundcolor=\color{backcolour},   
    commentstyle=\color{codegreen},
    keywordstyle=\color{codeblue},
    numberstyle=\tiny\color{codegray},
    stringstyle=\color{codeblue},
    basicstyle=\ttfamily\footnotesize,
    breakatwhitespace=false,         
    breaklines=true,                 
    captionpos=b,                    
    keepspaces=true,                 
    numbers=left,                    
    numbersep=5pt,                  
    showspaces=false,                
    showstringspaces=false,
    showtabs=false,                  
    tabsize=2
}
\usepackage{url}

\usepackage{flushend}
\usepackage{balance}

\usepackage[misc]{ifsym}
\usepackage{bbding}

\usepackage{xspace}

\usepackage{xspace}

 \usepackage[markup=underlined]{changes}

\theoremstyle{definition}
\newtheorem{definition}{Definition}[section]
\newcommand{\sysname}{{\tt jTrans}\xspace}
\newcommand{\benchname}{{\tt BinaryCorp}\xspace}

\begin{document}

\title{\sysname: Jump-Aware Transformer for Binary Code Similarity Detection} 

\author{Hao Wang}
\authornote{Both authors contributed equally to this research.}
\authornote{Institute for Network Sciences and Cyberspace, Tsinghua University}
\affiliation{%
  \institution{Tsinghua University, BNRist}
  \city{Beijing}
  \country{China}
}
\email{hao-wang20@mails.tsinghua.edu.cn}

\author{Wenjie Qu}
\authornotemark[1]
\affiliation{%
  \institution{Huazhong University of Science and Technology}
  \city{Wuhan}
  \country{China}
}
\email{wenjiequ@hust.edu.cn}

\author{Gilad Katz}
\affiliation{%
  \institution{Ben-Gurion University of the Negev}
  \city{Be'er Sheva}
  \country{Israel}
}
\email{giladkz@bgu.ac.il}

\author{Wenyu Zhu}
\affiliation{
  \institution{Tsinghua University, BNRist}
  \city{Beijing}
  \country{China}
}
\email{zhuwy19@mails.tsinghua.edu.cn}

\author{Zeyu Gao}
\affiliation{
  \institution{University of Science and Technology of China, Hefei, China}
  \city{Hefei}
  \country{China}
}
\email{zeyugao@mail.ustc.edu.cn}

\author{Han Qiu}
\authornotemark[2]
\affiliation{
  \institution{Tsinghua University}
  \city{Beijing}
  \country{China}
}
\email{qiuhan@tsinghua.edu.cn}

\author{Jianwei Zhuge}
\authornotemark[2]
\affiliation{
  \institution{Tsinghua Universityy, BNRist}
  \city{Beijing}
  \country{China}
}
\email{zhugejw@tsinghua.edu.cn}

\author{Chao Zhang}
\authornotemark[2]
\authornote{Zhongguancun Lab}
\authornote{Corresponding author}
\affiliation{
  \institution{Tsinghua University, BNRist}
  \city{Beijing}
  \country{China}
}
\email{chaoz@tsinghua.edu.cn}








\begin{abstract}
  Binary code similarity detection (BCSD) has important applications in various fields such as vulnerabilities detection, software component analysis, and reverse engineering. Recent studies have shown that deep neural networks (DNNs) can comprehend instructions or control-flow graphs (CFG) of binary code and support BCSD. In this study, we propose a novel Transformer-based approach, namely \sysname, to learn representations of binary code. It is the first solution that embeds control flow information of binary code into Transformer-based language models, by using a novel jump-aware representation of the analyzed binaries and a newly-designed pre-training task. Additionally, we release to the community a newly-created large dataset of binaries, \benchname, which is the most diverse to date. Evaluation results show that \sysname outperforms state-of-the-art (SOTA) approaches on this more challenging dataset by 30.5\% (i.e., from 32.0\% to 62.5\%). In a real-world task of known vulnerability searching, \sysname achieves a recall that is 2X higher than existing SOTA baselines.
    
\end{abstract}



\maketitle


\section{Introduction} \label{sec:introduction}



Binary code similarity detection (BCSD), which can identify the degree of similarity between two binary code snippets, is a fundamental technique useful for a wide range of applications,
including 
known vulnerabilities discovery~\cite{david2014tracelet, pewny2014leveraging, pewny2015cross, eschweiler2016discovre, david2016statistical, feng2016scalable, huang2017binsequence,feng2017extracting,david2017similarity, gao2018vulseeker, xu2017neural, david2018firmup, shirani2018binarm, liu2018alphadiff}, 
malware detection~\cite{cesare2013control} and clustering~\cite{hu2009large, hu2013mutantx, kim2019binary}, detection of software plagiarism~\cite{saebjornsen2009detecting, luo2014semantics, luo2017semantics},  patch analysis~\cite{hu2016cross,xu2017spain, kargen2017towards}, and software supply chain analysis~\cite{hemel2011finding}. Given the continuously expanding number of binary programs and the fact that binary analysis tasks are widespread, there is a clear need to develop BCSD solutions that are both more scalable and accurate.






Prior to the use of machine learning in the field, traditional BCSD solutions heavily relied on specific features of binary code, i.e., control flow graphs (CFGs) of functions, which capture the syntactic knowledge of programs.
Solutions such as BinDiff~\cite{bindiff}, BinHunt \cite{gao2008binhunt} and iBinHunt \cite{ming2012ibinhunt} employ graph-isomorphism techniques to calculate the similarity of two functions' CFGs. This approach, however, is both time-consuming and volatile, since CFGs may change based on compiler optimizations.
Studies such as BinGo~\cite{chandramohan2016bingo} and Esh~\cite{david2016statistical} achieve greater robustness to CFG changes by computing the similarities of CFG fragments. 
However, these approaches are based on manually crafted features, which have difficulty capturing the precise semantics of binary code. As a result, these solutions tend to have relatively low accuracy.


With the rapid development of machine learning techniques, most current state-of-the-art (SOTA) BCSD solutions are learning-based.
In general, these solutions embed target binary code (e.g., functions) into vectors, and compute functions' similarity in the vector space.
Some solutions, e.g., Asm2Vec~\cite{ding2019asm2vec} and SAFE~\cite{massarelli2019safe}, model assembly language (of machine code) using language models inspired by natural language processing (NLP). 
Other studies use graph neural networks (GNNs) to learn the representation of  CFGs and calculate their similarity~\cite{xu2017neural}. Some studies combine both approaches, and learn representations of basic blocks by NLP techniques and further process basic block features in a CFG by GNN, e.g., ~\cite{yu2020order,massarelli2019investigating}. Despite their improved performance, existing methods have several limitations. 

First, NLP-based modeling of assembly language only considers the sequential order of instructions and the relationships among them; information regarding the program's actual execution (e.g., control flows) is not considered. As a result, methods that rely solely on NLP will lack semantic understanding of the analyzed binaries, and will also not be adapt well to possibly significant changes in the code which are the result of compiler optimization.

Secondly, relying solely on CFGs misses semantics of instructions in each basic block. 
Genius~\cite{feng2016scalable} and Gemini~\cite{xu2017neural} propose to expand the CFG with manually extracted features (e.g., number of instructions).
However, such features are still insufficient to fully capture the code semantics.
Furthermore, these solutions generally use GNN to process CFGs, which only captures the structural information. GNNs are also generally known to be relatively difficult to train and apply in parallel, which limits their real-world application.

Thirdly, the datasets on which existing solutions are trained and evaluated are not sufficiently large and/or diverse. Due to the lack of a common large benchmark, each study creates its own dataset, often from small repositories such as GNUtils, coreutils, and openssl.
These small datasets have similar code patterns, and therefore lack diversity, which in turn can lead to over-fitted models and a false impression of high performance. 
Furthermore, the evaluations of existing solutions often does not reflect real-world use cases. 
The majority of studies did not conduct experiments on a large pool of candidate functions, which are common in the real world. Under more realistic conditions, the performance of many SOTA solutions drops significantly, as we show in our experimental results in Section \ref{evaluation}.

In this paper, we present \sysname, a novel Transformer-based model designed to address the aforementioned problems and support real-world binary similarity detection.
We combine NLP models, which capture the semantics of instructions, together with CFGs, which capture the control flow information, to infer the representation of binary code.
Since previous work~\cite{yu2020order} has shown that a simple combination of NLP-based and GNN-based features does not yield optimal results,
we propose to fuse the control-flow information into the Transformer architecture.
To the best of our knowledge, we are the first to do so.

We modify the Transformer to capture 
control-flow information, by sharing parameters between token embeddings and position embeddings for each jump target of instructions.
We first use unsupervised learning tasks to pre-train \sysname to learn the semantics of instructions and control-flow information of binary functions. Next, we fine-tune the pre-trained \sysname to match semantically similar functions.  
Note that our method is able to combine features from each basic block using language models without relying on GNNs to traverse the corresponding CFG.

In addition to our novel approach, we present a large and diversified dataset, \benchname, extracted from ArchLinux's official repositories~\cite{archlinux} and Arch User Repository (AUR) \cite{aur}. Our newly-created dataset enables us to mitigate the over-fitting and lack of diversity that characterize existing datasets. We automatically collect all the c/c++ projects from the repositories, 
which contain the majority of popular open-source softwares, and build them with different compiler optimizations to yield different binaries.
To the best of our knowledge, ours is the largest and most diversified binary program dataset for BCSD tasks to date.
 
We implement a prototype of \sysname and evaluate it on real-world BCSD problems, where we show that \sysname significantly outperforms SOTA solutions, including Gemini\cite{xu2017neural}, SAFE~\cite{massarelli2019safe}, Asm2Vec~\cite{ding2019asm2vec}, GraphEmb~\cite{massarelli2019investigating} and OrderMatters~\citep{yu2020order}. 
When using the full-sized \benchname 
in the task of finding the matching function in pools that have 10,000 functions, which is close to real world scenarios,
\sysname ranks the correct matching function the highest similarity score (denoted as Recall@1) with a probability of $62.5\%$ on average, while the best of SOTA solutions only achieves $32.0\%$.
For the less realistic (and easier) scenario of pools with 32 functions, our approach outperforms its closest competitor by 10.6\% (from $84.3\%$ to $94.9\%$) for the same Recall@1 metric. 
Furthermore, when evaluated on a real-world vulnerability searching task, \sysname achieves a recall score that is 2X higher than the SOTA baselines.

In summary, our study offers the following contributions:
\begin{itemize}
\item We propose a novel jump-aware Transformer-based model, \sysname, which is the first solution to embed control-flow information into Transformer. Our approach is able to learn binary code representations and support real world BCSD. We release the code of \sysname at \url{https://github.com/vul337/jTrans}.
\item We create a new large-scale, well-formed and diversified dataset, \benchname, for the task of BCSD. To the best of our knowledge, \benchname is the most diverse to date, and it can significantly mitigate the overfitting issues of previous benchmarks. 

\item  We conduct extensive experiments, and show that our model can significantly outperform SOTA approaches. 
\end{itemize}

\section{Problem Definition}
    
    


BCSD is a basic task to calculate the similarity of two binary functions.
It can be used in three types of scenarios as discussed in~\citep{haq2019survey}, 
including (1) {\em One-to-one (OO)}, where the similarity score of one \textit{source} function to one \textit{target} is returned; 
(2) {\em One-to-many (OM)}, where  a pool of {\em target} functions 
will be sorted based on their similarity scores to one \textit{source} function;
(3) {\em Many-to-many (MM)}, where a pool of functions will be divided into groups based on similarity.

Without loss of generality, we focus on OM tasks in this study. Note that, we can reduce OM problems to OO problems by setting the size of target functions to 1. We could also extend OM problems to MM, by taking each function in the pool as the {\em source} function and solving multiple OM problems.
To make the presentation clear, we give a formal definition of the problems as below:

\begin{figure}
  \centering
  \setlength{\abovecaptionskip}{2mm}
  \includegraphics[width=1\linewidth]{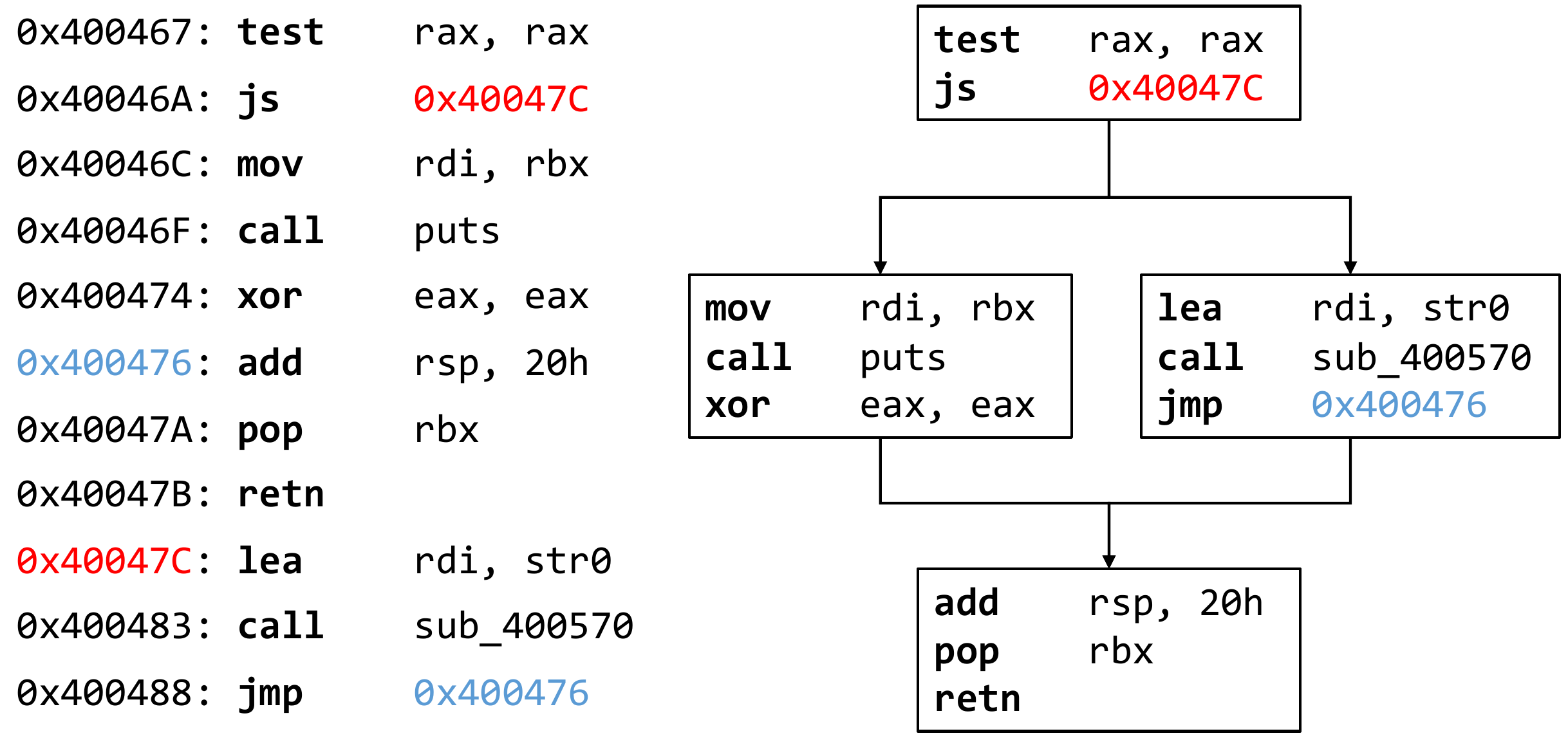}
  \caption{An example control flow of a binary function.
   The left part is linear layout assembly code with jump addresses, and the right part is the corresponding control-flow graph.}
   \label{fig:motivation_example}
  \vspace{-3mm}
\end{figure}

\begin{figure}
\begin{definition}[Function]
  In this study, we refer a function to a set of ordered instructions in binary programs, which are compiled from a source code function (maybe with inlined functions).
  It therefore has specific semantics and internal constraints. Especially, a function has a control-flow graph representing its control-flow information, as shown in Figure~\ref{fig:motivation_example}.
\end{definition}
\vspace{-7mm}
\end{figure}

The goal of BCSD is developing a solution to calculate the similarity scores of two functions, where two functions compiled from two source code functions that are same or similar to some extent (e.g., one is a patched version of the other) should be similar.

\begin{figure}
\begin{definition}[BCSD Task]
  Given a \textit{source} $f_q$ and a pool of functions $P$, the binary similarity detection task is to 
  retrieve the top-k \textit{functions} $\{f_{1}, f_{2}, \dots, f_{k} | f_{i} \in P\}$ ranked by the similarity score. 
\end{definition}
\vspace{-6mm}
\end{figure}

\section{Related Work}


\subsection{Non-ML-based BCSD Approaches} 
Prior to applying machine learning,
traditional binary code similarity techniques include static and dynamic methods. 
Under the assumption that logically similar code shares similar run-time behavior, dynamic analysis methods measure binary code similarity by analyzing manually-crafted dynamic features. This type of solutions includes works such as BinDiff~\cite{dullien2005graph}, 
BinHunt~\cite{gao2008binhunt}, iBinHunt~\cite{ming2012ibinhunt}, and Genius~\cite{feng2016scalable}, which are based on the CG/CFG graph-isomorphism (GI) Theory~\cite{dullien2005graph, flake2004structural}.
These works compare the similarity of two binary functions using graph matching algorithms. ESH~\cite{david2016statistical} employs a theorem prover to determine whether two basic blocks are equivalent. This approach, however, is not applicable to the case of different compiler optimizations due to basic block splitting. 
BinGo~\cite{chandramohan2016bingo}, Blex~\cite{egele2014blanket} and  Multi-MH~\cite{pewny2015cross} use randomly sampled values to initialize the context of the function and then 
compare the similarity by collecting the I/O values.
The main shortcoming of these dynamic methods is that they are not suitable for large-scale binary code similarity detection. This is due to the fact that they are computational expensive and require long running time to analyze the whole binary code.

Static methods for BCSD are based on the identification of structural 
differences in binary code. Methods such as BinClone~\cite{farhadi2014binclone}, ILine~\cite{jang2013towards}, MutantX-S~\cite{hu2013mutantx}, BinSign~\cite{nouh2017binsign}, and Kam1n0~\cite{ding2016kam1n0} use categorized operands or instructions 
as static features for the computation of binary similarity. 
Tracelet~\cite{david2014tracelet} and BinSequence~\cite{huang2017binsequence} compares the similarity of two binary functions based on the editing distance between instruction sequences.
TEDEM~\cite{pewny2014leveraging} and XMATCH~\cite{feng2017extracting} compute binary similarity using graph/tree edit distance of the basic block expression trees. 
While static methods are more efficient than dynamic ones, they generally achieve lower accuracy, as they only capture the structural and syntactical information of the binary, and neglect the semantics and relationship between instructions.  

\begin{figure}[t]
  \centering
  \setlength{\abovecaptionskip}{2mm}
  \includegraphics[width=0.9\linewidth]{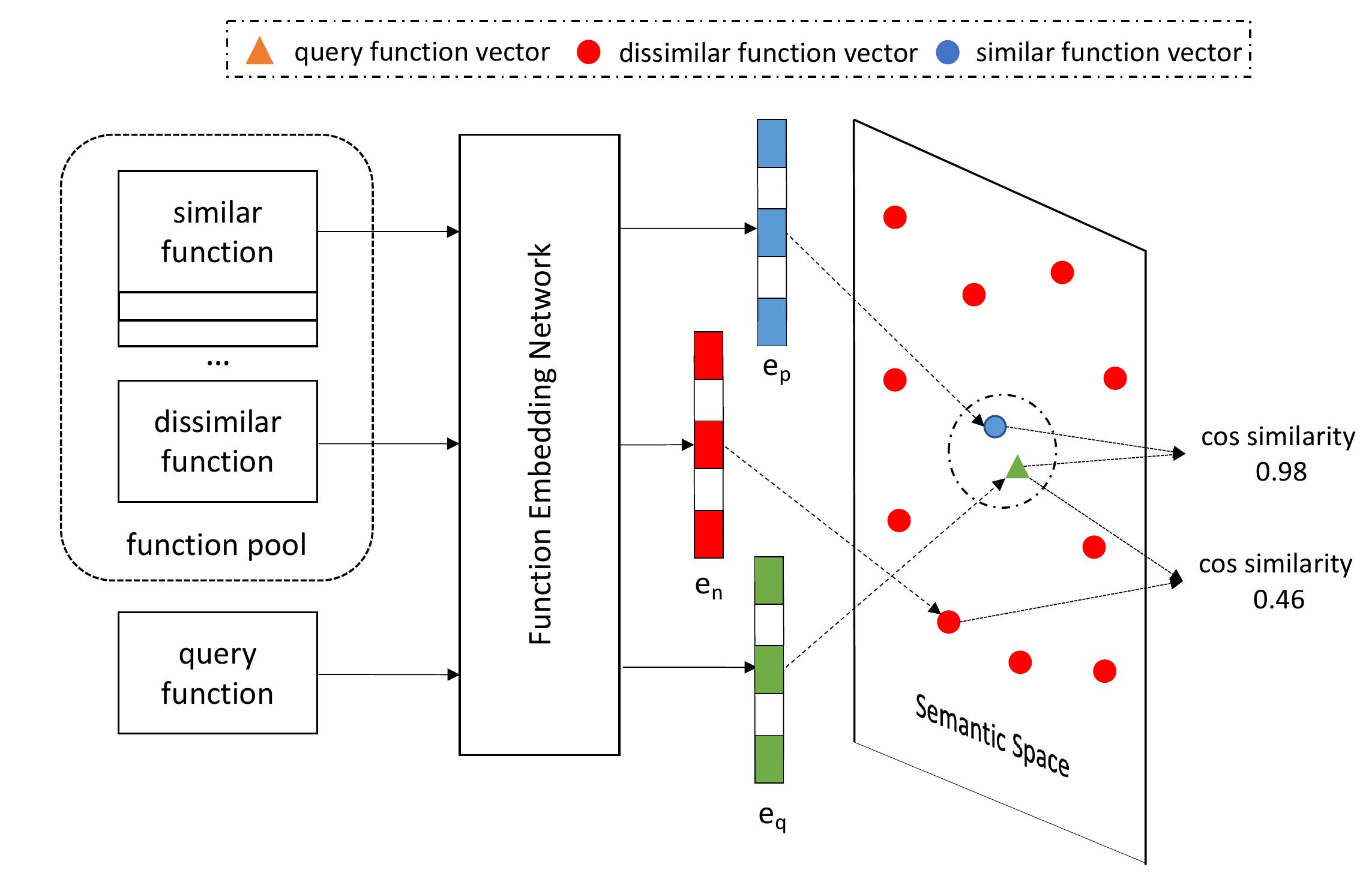}
  \caption{When using embeddings for binary code similarity detection, the query function and candidate functions in function pool are mapped into a semantic vector space. 
  The embeddings of query function, similar function and dissimilar function denote as $e_q, e_p \text{ and } e_n$, respectively.}\label{fig:learning_based_overview}
   \vspace{-3mm}
\end{figure}

\subsection{Learning-based BCSD Approaches}
\label{learning-based-binary-similarity-approaches}
The study of learning-based BCSD has been inspired by recent development in natural language processing (NLP)~\cite{mikolov2013efficient, le2014distributed, sutskever2014sequence}, which uses real-valued vectors called embeddings to encode semantic information of words and sentences. Building upon these techniques, previous studies~\cite{xu2017neural,liu2018alphadiff,gao2018vulseeker,redmond2018cross,zuo2018neural,ding2019asm2vec,massarelli2019safe,massarelli2019investigating,yu2020order,duan2020deepbindiff,yang2021codee} 
applied deep learning methods to binary similarity detection. Shared by many of these studies is the idea of \textit{embedding} binary functions into numerical vectors, and then using vector distance to approximate the similarity between different binary functions. As shown in Figure~\ref{fig:learning_based_overview}, these methods use deep learning training algorithms to make the vector distances of logically similar binary functions closer. 

Most learning-base methods use Siamese network~\cite{chopra2005learning}, 
which requires ground-truth mappings of equivalent binary functions to be trained. 
$\alpha \mathrm{Diff}$~\cite{liu2018alphadiff}, for example, learns binary function embeddings directly from the sequence of raw bytes using convolutional neural network (CNN)~\cite{krizhevsky2012imagenet}. 
INNEREYE~\cite{zuo2018neural} and RLZ2019~\cite{redmond2018cross} regard instructions as words and basic blocks as sentences, and use word2vec~\cite{mikolov2013efficient} and LSTM~\cite{hochreiter1997long} to learn basic block embeddings. 
SAFE~\cite{massarelli2019safe} uses a similar approach to learn the embeddings of binary functions, while 
Gemini~\citep{xu2017neural}, VulSeeker~\citep{gao2018vulseeker}, GraphEmb~\cite{massarelli2019investigating} and OrderMatters~\cite{yu2020order} use GNNs to build a graph embedding model for learning attributed control-flow graph (ACFG) of binary functions. Gemini and VulSeeker encode basic blocks with manually-selected features, while GraphEmb and OrderMatters use neural networks to learn the embeddings of basic blocks. 
Another approach is propsoed by DEEPBINDIFF~\cite{duan2020deepbindiff} and Codee~\cite{yang2021codee}, which uses neural networks to learn the embeddings of generated instruction sequences instead of embedding the ACFG of binary functions. 

Unsupervised learning (learing without labels) has also been explored in the field of BCSD. One representative solution is Asm2Vec~\cite{ding2019asm2vec}. This approach also generates instruction sequences with CFG, but does not rely on the ground truth mappings of equivalent binary functions. Asm2Vec uses an unsupervised algorithm to learn the embedding of binary functions. 
However, its performance is not as good as the state-of-the-art supervised learning methods. 

Overall, learning-based approaches are suitable for large-scale binary code similarity detection, as binary code functions can be transformed into vectors. Then similarity can be computed using vector distance, which is computationally efficient. However, existing techniques have limitations.
Some approaches~\cite{xu2017neural,gao2018vulseeker} ignore the semantics of instructions and basic blocks, as they only use manually-selected features to represent basic blocks.
Other approaches~\cite{liu2018alphadiff,redmond2018cross,zuo2018neural,massarelli2019safe, ding2019asm2vec,duan2020deepbindiff,yang2021codee} neglect some or all of the structural information of binary functions, as they do not use the control-flow information, or generate instruction sequences with CFG using random walk.
Finally, methods such as \cite{massarelli2019investigating, yu2020order} learn basic block embeddings, and use GNN to learn the embedding of attributed control-flow graph (ACFG) of binary functions. While these approaches are effective in some scenarios, they neglect the co-occurrence between inter-basic block instructions.
\section{Methodology}


\begin{figure*}[!ht]
  \centering
  \setlength{\abovecaptionskip}{2mm}
  \includegraphics[width=0.85\linewidth]{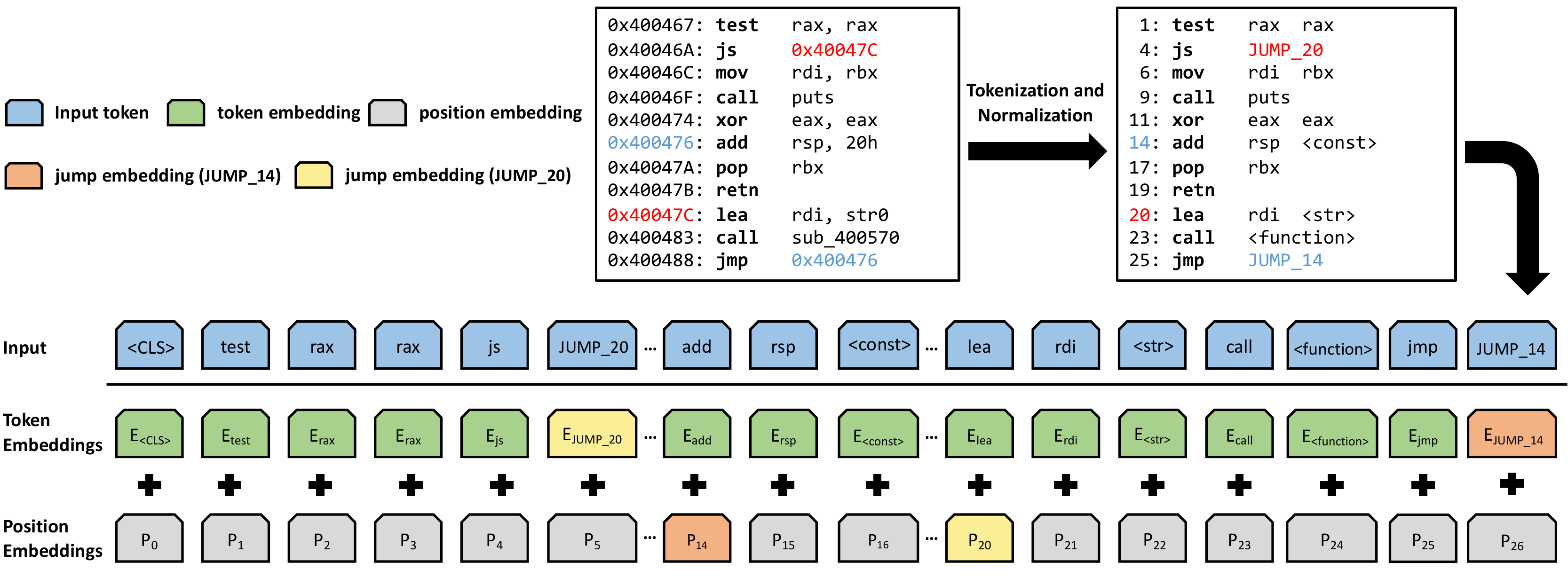}
  \caption{Input representation of \sysname. 
  The raw assembly code is first tokenized and normalized.
  Then, each token is converted to a {\em token embedding} and a {\em position embedding}, while its final input embedding is the sum of these two.
  For each jump pair, its source token's embedding (e.g., $\mathbf{E_\mathtt{JUMP\_14}}$), also called jump embedding, shares parameters with its target token's position embedding (e.g., $\mathbf{P_{\mathtt{14}}})$.
}
  \label{fig:input_representation}
\end{figure*}

\subsection{Overview}


To address the challenge we discussed in Section~\ref{sec:introduction}, we propose a novel model, \sysname, for automatically learning the instruction semantics and control-flow information of binary programs. 
\sysname is based on the Transformer-Encoder architecture~\cite{vaswani2017attention}, and consists of several significant changes designed to make it more effective for the challenging domain of binary analysis. 

The first change we propose to the Transformer architecture is designed to enable \sysname to better capture the code's jump relationships, i.e., the control-flow information. To this end, we first preprocess the assembly code of the input binary so that it contains the program's jump relationships. Next, we modify the embedding of the individual input tokens of the Transformer so that the origin and destination locations of the jumps are ``semantically'' similar.

The second change we propose relates to the training of our proposed model. In view of the similarity between natural language and programs in terms of data flow, we chose to use the commonly used effective Transformer training approach of Masked Language Model (MLM) ~\cite{devlin2018bert}. MLM-based training requires the model to predict the content of masked tokens based on the content of their neighbors, thus forcing the model to develop a contextual understanding of the relationships among instructions. 

Furthermore, to encourage the model to learn the manner by which jumps are incorporated in the code, we propose a novel auxiliary training task that requires the model to predict the target of a jump instruction. This task, which we call Jump Target Prediction (JTP), requires an in-depth understanding of the semantics of the code, and as shown in Section~\ref{sec:jump_design}, it contributes significantly to the performance of our model. 


\subsection{Binary Function Representation Model}
\sysname is based on the BERT architecture \cite{devlin2018bert}, which is the state-of-the-art pre-trained language model in many natural language processing (NLP) tasks. In \sysname, we follow the same general approach used by BERT for the modeling of texts, i.e., the creation of an embedding for each token (i.e., word), and the use of BERT's powerful attention mechanism to effectively model the binary code. 

However, binary codes are different from natural languages in several aspects. First, there are way too many vocabularies (e.g., constants and literals) in binary code. Second, there are jump instructions in binary code.
For a jump instruction, we denote its operand token as the {\em source token}, which specifies the address of the jump target instruction's address.
For simplicity, we denote the mnemonics token of the target instruction as the {\em target token}, and represent this \textit{jump pair} as <source token, target token>.

Therefore, we have to address two problems to apply BERT. 


\begin{itemize}
    \item \textbf{Out-of-vocabulary (OOV) tokens.} As in the field of NLP, we need to train \sysname on a fixed-size vocabulary that contains the most common tokens in the analyzed corpus. Tokens that are not included in the vocabulary need to be represented in a way that enables the Transformer to process them effectively. 
    
    \item \textbf{Modeling jump instructions.} 
    After preprocessing, the binary code has few information left for the source  and target token of a jump pair.
    BERT can hardly infer the connection between them.
    This problem is exacerbated by the possible large distance between the source and target, which makes contextual inference even more difficult.
    
    
\end{itemize}

We propose to address these challenges as follows. 

\subsubsection{Preprocessing instructions.}
To mitigate the OOV problem, we use the state-of-the-art disassembly tool IDA Pro 7.5 \cite{idapro} to analyze the input binary programs and yield sequences of assembly instructions. We then apply the following tokenization strategies to normalize  the assembly code and reduce its vocabulary size:
\begin{enumerate}
    \item We use the mnemonics and operands as tokens. 
    \item We replace the string literals with a special token \texttt{<str>}.
    \item We replace the constant values with a special token \texttt{<const>}.
    \item We keep the external function calls' names and labels as tokens, and replace the internal function calls' names as \texttt{<function>}\footnote{External function calls reflect interfaces between models and will not change frequently between different versions of binaries, but internal function calls are not}.
    \item For each jump pair, we replace its source token (which was absolute or relative address of the jump target) with a token \texttt{JUMP\_XXX}, where {\tt XXX} is the order of the target token of this jump pair, e.g., 20 and 14 in Figure~\ref{fig:input_representation}.
    In this way, we can remove the impact of random base addresses of binaries.
\end{enumerate}


\subsubsection{Modeling jump instructions.}
We now address the challenge of representing jump instructions in a manner that can enable \sysname to better contextualize their bipartite nature (and capture the entire control-flow information as a whole). We chose to use the \textit{positional encodings}, which are an integral part of the Transformer architecture. These encodings enable the model to determine the distance between tokens. The implicit logic of this representation is that larger distances between tokens generally indicate weaker mutual influence. Jump instructions, however, bind together areas in the code that may be far apart. Therefore we modify the positional encoding mechanism to reflect the effect of jump instructions.

Our changes to the positional encodings are designed to reflect the fact that the source and target of the jump instructions are not only as close as two consecutive tokens (due to order of execution), but also that they have a strong contextual connection. We achieve this goal through \textit{parameter sharing}: for each jump pair, the source token's embedding (see $E_{\texttt{JUMP\_14}}$ in Figure~\ref{fig:input_representation}) is used as the positional encoding of the target token (see $P_{14}$). 

This representation achieves two important goals. First, the shared embedding enables the Transformer to identify the contextual connection between the source and target tokens. Secondly, this strong contextual connection is maintained throughout the training process, since the shared parameters are updated for both tokens simultaneously. It is worth noting that we only focus on direct jump instructions in \sysname. We hypothesize that control flow information brought by indirect jumps will further improve the performance of \sysname. However, recognizing targets of such jumps is a well-known open challenge and beyond the scope of our current work. If a solution for recognizing indirect jump targets is proposed in the future, we can embed the operand of the indirect jump with the fusion of the positional embeddings of all jump targets.

\subsubsection{The Rationale of Our Proposed Approach} \label{sec:jTrans-for-function-representation}

By sharing the parameters between the source and target tokens of the jump pair, we create a high degree of similarity in their representation. As a result, whenever the attention mechanisms that power \sysname assign a high attention weight to one of these tokens (i.e., determine that it is important to the understanding/analysis of the binary), they will \textit{automatically} also assign high attention to their partner. This representation therefore ensures that both parts of the jump instruction---and the instructions near them in the code---will be included in the inference process.

We now provide a formal analysis of the jump pair's token similarity, and demonstrate that the similarity within this pair is higher than any of them has with any other token. For a given binary function $\mathbf{f} = [x_{1},\cdots, x_{n}]$, $x_i$ is the $i$-th token of $\mathbf{f}$. 
All the tokens will be converted into mixed embedding vectors $\{E(x_{1}),\cdots, E(x_{n})\}$ before being fed into \sysname, 
where each embedding $E(x_{i})$ can be represented as a summation of token embeddings $E_{x_i}$ and position embeddings $P_{i}$.
We apply the multi-head self-attention mechanism~\cite{vaswani2017attention} to the mixed embedding vectors $\{E(x_{1}),\cdots, E(x_{n})\}$. 
We denote the embedding of the $m$-th layer as $E_m = [E_{m}(x_{1}),\cdots, E_{m}(x_{n})]$, we first project the $m$-th embedding to $Q_m$, $K_m$ and $V_m$, respectively. 
Then we used the scaled dot-product attention to get the attention matrix $\text{Attention}(Q_m, K_m, V_m)$.
\begin{equation}
\small
  \begin{aligned}
  Q_m = E_m \times W^Q_m , K_m = E_m \times W^K_m , V_m = E_m \times W^V_m \\ 
  \text{Attention}(Q_m,K_m, V_m) = \text{Softmax}(\frac{Q_m K^T_m}{\sqrt{d_k}}) \cdot V_m \\
  \end{aligned}
\end{equation}

The $W^Q_m \in \mathbb{R}^{d_{\text{model}} \times d_k}$, $W^K_m \in \mathbb{R}^{d_{\text{model}} \times d_k}$, $W^V_m \in \mathbb{R}^{d_{\text{model}} \times d_k}$ are affine transformation matrices of the $m$-th layer, $d_{\text{model}}$ is the dimension of the embedding vector. $\text{Softmax}(\frac{Q_m K^T_m}{\sqrt{d_k}})$ is the attention weight matrix. We denote the updated embedding by head $h$ as 
\begin{equation}
\small
  \begin{aligned}
  E^h_{m+1} = \text{Attention}(Q^h_m, K^h_m, V^h_m) 
  \end{aligned}
\end{equation}
Assume we have $H$ attention head, we get updated embedding $E_{m+1}$ as follows, 
$W^O_m \in \mathbb{R}^{d_kH \times d_{\text{model}}}$ is the output transformation matrix of the $m$-th layer, $FFN_m$ is the feed-forward network of the $m$-th layer.
\begin{equation}
 \begin{aligned}
 E_{m+1} = FFN_m(\text{Concat}(E^1_m, \cdots, E^H_m) \times W^O_m) \\
 \end{aligned}
\end{equation}
The final output of jTrans is the last layer of the model. 
We produce the function embedding $E_f$ as follows, where $W^F \in \mathcal{R}^{d_{\text{model}} \times d_{f}}$ is the output transformation matrix, $d_{f}$ is the dimension of the function embedding, $E_N(C)$ represents the embedding of \texttt{<CLS>}. 

\begin{equation}\label{eq:function-embedding}
\small
  \begin{aligned}
    E_f = tanh(E_{N}(C)) \cdot W^F \\
  \end{aligned}
\end{equation}

Next, we present how \sysname deliver the control-flow information of the program.
Consider three tokens $i$, $j$, $l$ in the given function, the corresponding embeddings are $E_{i}$, $E_{j}$ and $E_{l}$. 
Assume we have a jump relationship between $i$ and $j$, $i$ is the source token and $j$ is the target token. 
And $l$ is any other token in the function. Denote the attention weight of token $i$ to token $j$ as $A_{ij} = \frac{Q_i K_j^T}{\sqrt{d_k}}$.
We prove that the mathematical expectation of $A_{ij}$ minus $A_{il}$ is positive. Which can be formulated as follows
\begin{equation}
\small
  \mathbb{E}[A_{ij} - A_{il}] > 0
\end{equation}
This equation shows that $i$ generally pays more attention to $j$ than $l$. This is the internal explanation of jump embeddings.
The detailed proof is in Appendix~\ref{proof_jump_embedding}. 
\subsection{Pre-training jTrans}

The BERT architecture, on which \sysname is based, uses two unsupervised learning tasks for pre-training. The first task is masked language models (MLM), where BERT is tasked with reconstructing randomly-masked tokens. The second unsupervised learning task is designed to hone BERT's contextual capabilities by requiring it to determine whether two sentences are consecutive. We build upon BERT's overall training process, while performing domain-specific adaptations: we preserve the MLM task, but replace the second task with one we call \textit{jump task prediction} (JTP). As shown in Section \ref{subsubsec:JTP}, the goal of the JTP task is to improve \sysname's contextual understanding of jump instructions.

\subsubsection{The Masked Language Model Task}
Our MLM task closely follows the one proposed in \cite{devlin2018bert}, and \sysname uses BERT's masking procedures: 80\% of our randomly-selected tokens are replaced by the \textit{mask} token (indicating that they need to be reconstructed), 10\% are replaced by other random tokens, and 10\% are unchanged. Following the notation in Section~\ref{sec:jTrans-for-function-representation}, we define the function $\mathbf{f} = [x_{1},\cdots, x_{n}]$, where $x_i$ is the $i$-th token of $\mathbf{f}$, and $n$ is the number of tokens. We first select a random set of positions for $\mathbf{f}$ to mask out~(i.e., $\mathbf{m^{x}}$). 

\begin{equation}
\small
  \begin{aligned}
  \mathbf{f}^{\text{mlm}} = \text{REPLACE}(\mathbf{f},\mathbf{m^{x}},\texttt{<MASK>})\\
  \end{aligned}
\end{equation}
Based on these definitions, the MLM objective of reconstructing the masked tokens can be formulated as follows:
\begin{equation}
\small
  \min_{\theta} \mathcal{L}_{MLM}(\theta) = \sum_{i \in \mathbf{m_{x}}} -\log P(x_i|\mathbf{f}^{\text{mlm}})
\end{equation}
where $\mathbf{m^{x}}$ contains the indices of the masked tokens.  

\begin{figure}[t]
  \centering
  \setlength{\abovecaptionskip}{2mm}
  \includegraphics[width=\linewidth]{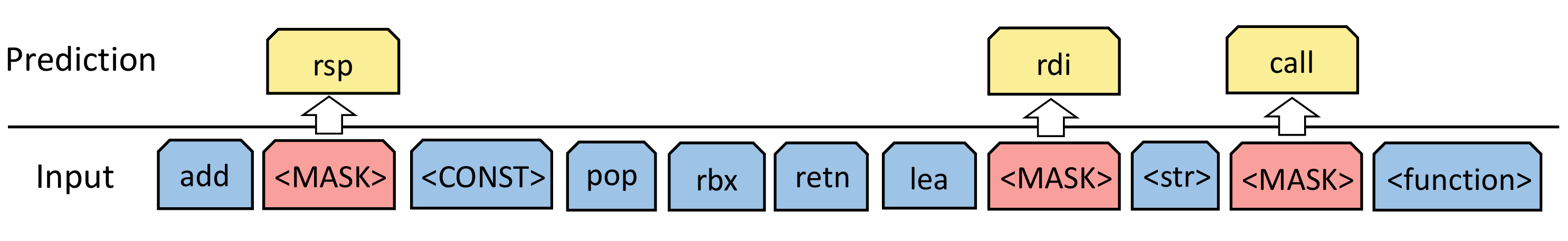}
  \caption{Masked Language Model~(MLM)}\label{MLM}
\end{figure}

An example of the masking process is presented in Figure~\ref{MLM}. We make the \texttt{rsp}, \texttt{rdi} and \texttt{call} tokens, and task \sysname with reconstructing them. To succeed, our model must learn the basic assembly syntax and its contextual information. Successfully reconstructing the \texttt{rdi} token, for example, requires that the model learns the \textit{calling convention} of the function, while the \texttt{rsp} token requires an understanding of continuous execution.

\subsubsection{Jump Target Prediction} \label{subsubsec:JTP}

The JTP task is defined as follows: given a randomly selected jump source token, our model is required to predict its corresponding target token. This task, which is difficult even for human experts, requires our model to develop a deep understanding of the CFG. This in turn leads to improved performance for \sysname, as we later show in Section~\ref{sec:jump_design}.
JTP is carried out by first selecting a random subset of the available jump source tokens. These tokens are then replaced with the token \texttt{<LOC>}.
\begin{equation}
\small
  \begin{aligned}
  \mathbf{f}^{\text{jtp}} = \text{REPLACE}(\mathbf{f},\mathbf{l^{x}},\texttt{<LOC>})\\
  \end{aligned}
\end{equation}
Where $\mathbf{l^{x}}$ is the set of positions for jump symbols. JTP's objective function can be formulated as follows:
\begin{equation}
\small
  \min_{\theta} \mathcal{L}_{JTP}(\theta) = \sum_{i \in \mathbf{l_{x}}} -\log P(x_i|\mathbf{f}^{\text{jtp}})
\end{equation}

\begin{figure}[t]
  \centering
  \setlength{\abovecaptionskip}{2mm}
  \includegraphics[width=\linewidth]{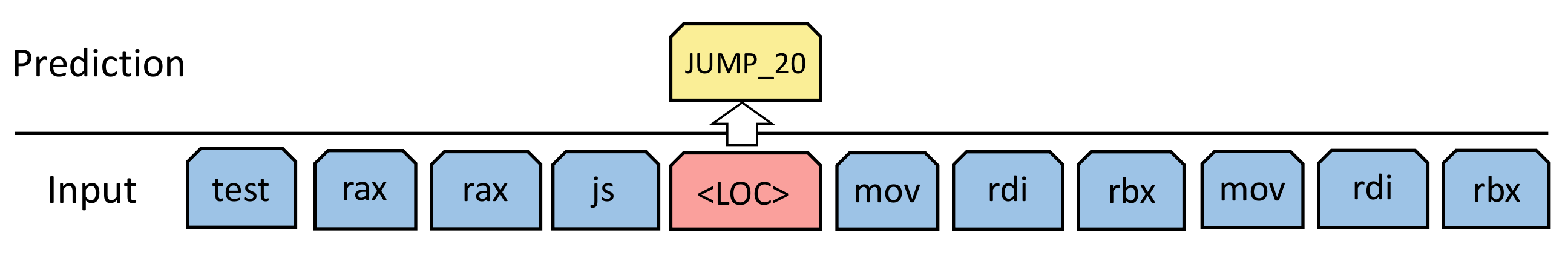}
  \caption{Jump Target Prediction~(JTP)}\label{JTP}
\end{figure}

An example of the JTP task is presented in Figure~\ref{JTP}, where we replace the \texttt{JUMP\_20} token by \texttt{<LOC>} and the model is tasked with predicting the index. An analysis of \sysname's performance in the JTP task is presented in Table~\ref{tab:JTP}, which shows that our model achieves the accuracy of 92.9\%. Furthermore, an ablation study that evaluates JTP's contribution to \sysname's overall performance is presented in Section~\ref{sec:jump_design}. These results provide a clear indication that this training task improves \sysname's ability to learn the control flow of analyzed functions.

The \textit{overall loss function} of \sysname in the pre-training phase is the summation of the MLM and JTP objective functions:

\begin{equation}
\small
  \min_{\theta} \mathcal{L}_{P}(\theta) = \mathcal{L}_{MLM}(\theta) + \mathcal{L}_{JTP}(\theta)
\end{equation}

\subsection{Fine-tuning for Binary Similarity Detection}

Upon the completion of the unsupervised pre-training phase, we now fine-tune our model for the supervised learning task of function similarity detection. We aim to train \sysname to maximize the similarity between similar binary functions pairs, while minimizing the similarity for unrelated pairs. As shown in Section ~\ref{sec:jTrans-for-function-representation}, we use Equation~\ref{eq:function-embedding} to represent function  $\mathbf{f}$. Our chosen metric for calculating function similarity is Cosine similarity.

We use the following notations: let $\mathcal{F}$ and $\mathcal{G}$ be a set of binary functions, and the sets of similar functions (the ``ground truth''). For any query function $\mathbf{f} \in \mathcal{F}$, let $\mathbf{g}^{+} \in \mathcal{G}$ be a function that is similar to $f$ (e.g., compiled from the same source code). Furthermore, let $\mathbf{g}^{-} \in \mathcal{G}$ be an arbitrary function, unrelated to $\mathbf{f}$. We denote the embedding for function $\mathbf{f}$ as $E_{\mathbf{f}}$. Finally, we define $\mathcal{D}$ as the set of all generated triples $\left\langle \mathbf{f}, \mathbf{g}^+, \mathbf{g}^-\right\rangle$. 

The objective function for our fine-tuning process in trained using contrastive learning~\cite{hadsell2006dimensionality,schroff2015facenet} and is performed as follows:
\begin{equation}\label{eq:cl-loss}
\small
  \min_{\theta} \mathcal{L}_{F}(\theta) = \sum_{\langle \mathbf{f}, \mathbf{g}^+ , \mathbf{g}^-\rangle \in \mathcal{D}} \max \left(0, \varepsilon - \cos\left(E_{\mathbf{f}}, E_{\mathbf{g}^{+}}\right) + \cos\left(E_{\mathbf{f}}, E_{\mathbf{g}^{-}}\right)\right)
\end{equation}
where $\theta$ represents the parameters of the model, and $\varepsilon$ is a hyper-parameter usually chosen between 0 and 0.5~\cite{paszke2019pytorch}. 
After fine-tuning \sysname on $\mathcal{D}$, we can measure the similarity score of two functions $f_1, f_2$ by calculating the cosine similarity of their embeddings. Once the fine-tuning process is complete, we can measure the similarity score of two functions $f_1, f_2$ by calculating the cosine similarity of their embeddings.

\subsection{Large-Scale Dataset Construction}

We build our dataset for binary similarity detection based on the ArchLinux~\cite{archlinux} official repositories and Arch User Repository~\cite{aur}.
ArchLinux is a Linux distribution known for its large number of packages and rapid package updates. 
ArchLinux's~\cite{archlinux} official repositories contain tens of thousands of packages, including editor, instant messenger, HTTP server, web browser, compiler, graphics library, cryptographic library, etc. And Arch User Repository contains more than 77,000 packages uploaded and maintained by users, greatly enlarging the dataset.
Furthermore, ArchLinux provides a useful tool \texttt{makepkg} for developers to build their packages from source code. 
\texttt{makepkg} can compile the specified package from the source by parsing the \texttt{PKGBUILD} file, which contains the required dependencies and compilation helper functions. 
Binary code similarity task requires a large number of labeled data, thus we use these infrastructures to construct our dataset.

\subsubsection{Projects Filtering}
For compilation compatibility reasons, we choose the c/c++ project in the pipeline to build the datasets. If the \texttt{build} function in \texttt{PKGBUILD} file contains the call of \texttt{cmake}, \texttt{make}, \texttt{gcc} and \texttt{g++}, then it is very likely to be a C/C++ project. On the other hand, if the variable \texttt{depend} in \texttt{PKGBUILD} file contains \texttt{rustc, go, jvm}, then it is not likely to be a C/C++ project, we can remove it before compiling.
\subsubsection{Compilation pipeline}
 In our pipeline, we wish to automatically specify any optimization level we want each time. The toolchain of some projects do not consume environment variables \texttt{CFLAGS} and \texttt{CXXFLAGS}, making it impossible to change the optimization level easily. However, because most projects call the compiler by \texttt{CC} or \texttt{CXX}, we assign environment variables a self-modified version of \texttt{gcc, g++, clang, clang++}. The modified compiler changes the command line parameters that are related to the optimization level to expected compilation parameters. Also, it appends the expected compilation arguments to the original parameters. We use these two ways to ensure the compilation is done with the  expected optimization level. 
\subsubsection{Label Collection}
To collect labels, we need to first obtain unstripped binary and get the offset of functions. We found that many real-world projects call \texttt{strip} during compilation, therefore only specifying parameters in \texttt{PKGBUILD} doesn't solve this problem. We replaced the \texttt{strip} with its modified version. It will not strip the symbol table regardless of passed-in parameters. 

\section{Experimental Setup}

\begin{table}[t]
\footnotesize
  \setlength{\abovecaptionskip}{2mm}
  \caption{Statistics on the number of projects, binaries and functions of the datasets. Project refers to binaries compiled from the same source code. }
  \begin{tabular}{cccccc}
  \Xhline{1.5pt}
  Datasets              & \# Projects & \# Binaries & \# Functions  \\ \hline
  GNUtils               & 20          & 100         & 161,202           \\
  Coreutils             & 115         & 575         & 76,174           \\ \hline
  \benchname-3M~Train  & 1,612       & 8,357       & 3,126,367       \\
  \benchname-3M~Test   & 364         & 1,908       & 444,574         \\ \hline
  \benchname-26M~Train & 7,845       & 38,455      & 21,085,338    \\
  \benchname-26M~Test  & 1,974       & 9,675       & 4,791,673       \\ \hline
  \Xhline{1.5pt}
  \end{tabular}
  \label{asm_stat}
\end{table}

\subsection{The BinaryCorp Dataset}
\label{sectionDataset}

We now present \benchname, the dataset we created to evaluate large-scale binary similarity detection. 
\benchname consists of a large number of binaries produced by automatic compilation pipeline, where---based on the official ArchLinux packages and Arch User Repository---we use \texttt{gcc} and \texttt{g++} to compile 48,130 binary programs with different optimization levels and follow the approach proposed in SAFE\cite{massarelli2019safe} to filter duplicate functions.

The statistics of our datasets are shown in Table~\ref{asm_stat}. While many previous works use Coreutils and GNUtils as their dataset, Table~\ref{asm_stat} clearly shows that \benchname-26M operates at a different scale: while our newly-created dataset has approximately 26 million functions compared to GNUtils' 161,202 and Coreutils 76,174. \textit{\benchname-26M is therefore more than 160 times the size of GNUtils and more than 339 the size of Coreutils.}

The size of our new dataset prevents the use of some of the existing methods, due to their insufficient scalability. We therefore also provide a smaller dataset, named \benchname-3M, which contains 10,265 binary programs and about 3.6 million functions. The number of functions in our smaller dataset is about 22 times that of GNUtils, and 47 times that of Coreutils.

BinKit\cite{kim2020revisiting} is the largest binary dataset, which consists 36,256,322 functions. However, BinKit used 1,352 different compile options to generate 243k binaries from only 51 GNU packages, therefore having too many similar functions in the dataset. We, on the other hand, only use 5 different compile options to compile nearly 10,000 projects. While our number of binaries is smaller, our dataset is more diversified than BinKit, in terms of developers, project size, coding style and application scenarios. We argue that our newly generated dataset offers a more diverse---and therefore more realistic---basis for learning and evaluation. Using our datasets, we can evaluate the scalability and efficacy of \sysname (and the other baselines) on a new and larger scale. As mentioned in our contributions in Section~\ref{sec:introduction}, we will make our datasets and trained models available to the community.


\subsection{Baselines}
\label{sectionBaseline}

We compare \sysname to six top-performing baselines:

    \textbf{Genius~\cite{feng2016scalable}.} The baseline is a non-deep learning approach. Genius extracts raw features in the form of an attributed control flow graph and uses locality sensitive hashing (LSH) to generate numeric vectors for vulnerability search. We implemented this baseline based on its official code\footnote{https://github.com/Yunlongs/Genius}.
    
    \textbf{Gemini~\cite{xu2017neural}.} This baseline extracts manually crafted features for each basic block, and uses GNN to learn the CFG representation of the analyzed function. We implemented this approach based on its official Tensorflow code\footnote{https://github.com/xiaojunxu/dnn-binary-code-similarity}, and used its default parameter settings throughout our evaluation.
    
    \textbf{SAFE~\cite{massarelli2019safe}.} This baseline employs an RNN architecture with attention mechanisms to generate a representation of the analyzed function, it receives the assembly instructions as input.
    We implemented this baseline based on its official Pytorch code\footnote{https://github.com/facebookresearch/SAFEtorch}, and default parameter settings. 
    
    \textbf{Asm2Vec~\cite{ding2019asm2vec}.} The method uses random walks on the CFG to sample instruction sequences, and then uses the PV-DM model to jointly learn the embedding of the function and instruction tokens. 
    This approach is not open source, 
    and we therefore used an unofficial implementation \footnote{https://github.com/oalieno/asm2vec-pytorch}. 
    We used its default parameter settings.
    
    \textbf{GraphEmb~\cite{massarelli2019investigating}.} This baseline uses word2vec~\cite{mikolov2013efficient} to learn the embeddings of the  instruction tokens.
    Next, it uses a RNN to generate independent embeddings for each basic block, and finally uses structure2vec~\cite{dai2016discriminative} to combine the embeddings and generate representation of the analyzed function.
    To make this baseline scalable to datasets as large as \benchname-26M,
    we re-implemented the author's original Tensorflow source code\footnote{https://github.com/lucamassarelli/Unsupervised-Features-Learning-For-Binary-Similarity} using Pytorch. 
    
    \textbf{OrderMatters~\cite{yu2020order}.} This method combines two types of embeddings. 
    The first embedding type uses BERT to create an embedding for each basic block, with all these embeddings then combined using a GNN to generate the final representation. 
    The second type of embeddings is obtained by applying a CNN on the CFG. 
    The two embeddings are then concatenated. 
    This method is not open source,
    and its online blackbox API\footnote{https://github.com/binaryai/sdk} can not satisfy the need of this study. We implemented on our own using the reported hyperparameters.
    

\renewcommand\arraystretch{0.9}
\begin{table*}[h!]
\centering
\caption{Results of different binary similarity detection methods on \benchname-3M (Poolsize=32)}
\label{tab:bench-3M-32}
\scalebox{0.85}{
\begin{threeparttable}
\begin{tabular}{c|ccccccc|ccccccc}
\Xhline{1.5pt}
\hline
             & \multicolumn{7}{c|}{\textbf{MRR}}                             & \multicolumn{7}{c}{\textbf{Recall@1}}                               \\ \hline
\textbf{Models}       & \textbf{O0,O3} & \textbf{O1,O3} & \textbf{O2,O3} & \textbf{O0,Os} & \textbf{O1,Os} & \textbf{O2,Os} & \textbf{Average} & \textbf{O0,O3} & \textbf{O1,O3} & \textbf{O2,O3} & \textbf{O0,Os} & \textbf{O1,Os} & \textbf{O2,Os} & \textbf{Average}  \\ \hline
Gemini       & 0.388 & 0.580 & 0.750 & 0.455 & 0.546 & 0.614 & 0.556  & 0.238 & 0.457 & 0.669 & 0.302 & 0.414 & 0.450 & 0.422   \\
SAFE         & 0.826 & 0.917 & 0.958 & 0.854 & 0.927 & 0.927 & 0.902   & 0.729 & 0.869 & 0.933 & 0.766 & 0.879 & 0.880 & 0.843    \\
Asm2Vec      & 0.479 & 0.878 & 0.961 & 0.536 & 0.855 & 0.900 & 0.768   & 0.351 & 0.828 & 0.942 & 0.408 & 0.796 & 0.863 & 0.701   \\
GraphEmb     & 0.602 & 0.694 & 0.750 & 0.632 & 0.674 & 0.675 & 0.671   & 0.485 & 0.600 & 0.678 & 0.521 & 0.581 & 0.584 & 0.575          \\
OrderMatters-online\tnote{1} & 0.542  & 0.740 & 0.869 & 0.638 & 0.702 & 0.682   & 0.695 & 0.414 & 0.647 & 0.822 & 0.515 & 0.611 & 0.593 & 0.591   \\ 
OrderMatters & 0.601 & 0.838 & 0.933 & 0.701 & 0.812 & 0.800 & 0.777   & 0.450 & 0.763 & 0.905 & 0.566 & 0.724 & 0.715  & 0.687          \\ 
Genius & 0.377  &0.587  &0.868  &0.437  & 0.600 &0.627    &0.583  &0.243  &0.479  &0.830 &0.298   & 0.490 &0.526  & 0.478   \\ \hline
\sysname-Zero\tnote{2}  & 0.594 & 0.841 & 0.962 & 0.649 & 0.850 & 0.891 & 0.797 & 0.499 & 0.803  & 0.945 & 0.566 & 0.808 & 0.853 & 0.746    \\ 
\sysname       & \textbf{0.947} & \textbf{0.976} & \textbf{0.985} & \textbf{0.956} & \textbf{0.979} & \textbf{0.977} & \textbf{0.970} & \textbf{0.913} & \textbf{0.960} & \textbf{0.974} & \textbf{0.927} & \textbf{0.964} & \textbf{0.961} & \textbf{0.949} \\
\Xhline{1.5pt}
\end{tabular}
		\begin{tablenotes}
			\item[1] We evaluate OrderMatters with the API provided at https://github.com/binaryai/sdk.
			\item[2] \sysname-Zero represents \sysname without fine-tuning.
		\end{tablenotes}
	\end{threeparttable}
}
\end{table*}

\begin{table*}[h!]
\renewcommand\arraystretch{0.9}
\caption{Results of different binary similarity detection methods on \benchname-3M (Poolsize=10000)}
\centering
\scalebox{0.85}{
\begin{tabular}{c|ccccccc|ccccccc}
\Xhline{1.5pt}
\hline
             & \multicolumn{7}{c|}{\textbf{MRR}}                             & \multicolumn{7}{c}{\textbf{Recall@1}}                               \\ \hline
\textbf{Models}       & \textbf{O0,O3} & \textbf{O1,O3} & \textbf{O2,O3} & \textbf{O0,Os} & \textbf{O1,Os} & \textbf{O2,Os} & \textbf{Average} & \textbf{O0,O3} & \textbf{O1,O3} & \textbf{O2,O3} & \textbf{O0,Os} & \textbf{O1,Os} & \textbf{O2,Os} & \textbf{Average}  \\ \hline
Gemini         & 0.037 & 0.161 & 0.416 & 0.049 & 0.133 & 0.195 & 0.165   & 0.024 & 0.122 & 0.367 & 0.030 & 0.099 & 0.151 & 0.132    \\
SAFE         & 0.127 & 0.345 & 0.643 & 0.147 & 0.321 & 0.377 & 0.320 & 0.068 & 0.247 & 0.575 & 0.079 & 0.221 & 0.283 & 0.246  \\
Asm2Vec      & 0.072 & 0.449 & 0.669 & 0.083 & 0.409 & 0.510 & 0.366 & 0.046 & 0.367 & 0.589 & 0.052 & 0.332 & 0.426 & 0.302  \\
GraphEmb     & 0.087 & 0.217 & 0.486 & 0.110 & 0.195 & 0.222 & 0.219 & 0.050 & 0.154 & 0.447 & 0.063 & 0.135 & 0.166 & 0.169          \\ 
OrderMatters & 0.062 & 0.319 & 0.600 & 0.075 & 0.260 & 0.233 & 0.263 & 0.040 & 0.248 & 0.535 & 0.040 & 0.178 & 0.158 & 0.200          \\ 
Genius &0.041   &0.193  &0.596  &0.049  &0.186  &0.224    &0.214  &0.028  &0.153  &0.538  &0.032  &0.146  &0.180  & 0.179   \\ \hline
\sysname-Zero  & 0.137 & 0.490 & 0.693 & 0.182 & 0.472 & 0.510 & 0.414 & 0.088 & 0.412  & 0.622 & 0.122 & 0.393 & 0.430 & 0.340  \\ 
\sysname       & \textbf{0.475} & \textbf{0.663} & \textbf{0.731} & \textbf{0.539} & \textbf{0.665} & \textbf{0.664} & \textbf{0.623} & \textbf{0.376} & \textbf{0.580} & \textbf{0.661} & \textbf{0.443} & \textbf{0.586} & \textbf{0.585} & \textbf{0.571} \\ \hline
\Xhline{1.5pt}
\end{tabular}
}\label{tab:bench-3M-10000}
\end{table*}

\subsection{Evaluation Metrics}
Let there be a binary function pool $\mathcal{F}$, and its ground truth binary function pool $\mathcal{G}$.

\begin{equation}
\small
  \begin{aligned}
    &\mathcal{F}=\{f_{1}, f_{2}, ..., f_{i}, ..., f_{n}\} \\
    &\mathcal{G}=\{f_{1}^{gt}, f_{2}^{gt}, ..., f^{gt}_{i}, ..., f_{n}^{gt}\}
  \end{aligned}
\end{equation}

We denote a query function $f_{i} \in \mathcal{F}$ and its corresponding ground truth function $ f^{gt}_{i} \in \mathcal{G}$. 
In this study we address the binary similarity detection problem, and our goal is therefore to retrieve the top-k functions in function pool $\mathcal{G}$, which have the highest similarity to $f_{i}$. The returned functions are ranked by a similarity score,
$Rank_{f^{gt}_{i}}$ which denotes their position in the list of retrieved functions. The indicator function $\mathbb{I}$ 
is defined as below

\begin{equation}
\small
  \mathbb{I}(x)=
  \begin{cases}
  0, &  x = False\\
  1, &  x = True
  \end{cases}
\end{equation}

The retrieval performance can be evaluated using the following two metrics:

\begin{equation}
\small
  \begin{aligned}
  &\text{Recall}@k = \frac{1}{|\mathcal{F}|} \sum_{f_{i}\in \mathcal{F}}\mathbb{I}(\text{Rank}_{f^{gt}_{i}} \le k) \\
  & \text{MRR} = \frac{1}{|\mathcal{F}|} \sum_{f_{i} \in \mathcal{F}} \frac{1}{Rank_{f^{gt}_{i}}}
  \end{aligned}
\end{equation}

\section{Evaluation}\label{evaluation}


Our evaluation aims to answer the following questions.
\begin{itemize}
\item \textbf{RQ1:} How accurate is \sysname in BCSD tasks compared with other baselines? (\S\ref{subsec:RQ1})
\item \textbf{RQ2:} How well do \sysname and baselines perform on BCSD tasks of different function pool sizes?  (\S\ref{subsec:RQ2})
\item \textbf{RQ3:} How effective is \sysname at discovering known vulnerability? (\S\ref{subsec:RQ3})
\item \textbf{RQ4:} How effective is our jump-aware design? (\S\ref{sec:jump_design})
\item \textbf{RQ5:} How effective is the pre-training design? (\S\ref{subsec:RQ5})
\end{itemize}


Throughout our experiments, all binary files were initially stripped to prevent information leakage. We used IDA Pro to disassemble and extract the functions from the binary code in all of the experiments, thus ensuring a level playing field. For baselines that didn't use IDA Pro, we used their default disassemble frameworks for preprocessing, after extracting functions using IDA Pro. All the training and inference were run on a Linux server running Ubuntu 18.04 with Intel Xeon 96 core 3.0GHz CPU including hyperthreading, 768GB RAM and 8 Nvidia A100 GPUs. 
\subsection{Biniary Similarity Detection Performance} \label{subsec:RQ1}


We conduct our evaluation on our two datasets, \benchname-3M and \benchname-26M. Additionally, we use two function pool sizes---32 and 10,000---so that \sysname and the baselines can be evaluated in varying degrees of difficulty. It is important to note that we randomly assign entire projects to either the train or test sets of our experiments, because recent studies \cite{nucleus} have shown that randomly allocating binaries may result in information leakage. 

The results of our experiments are presented in Tables \ref{tab:bench-3M-32}--\ref{tab:bench-26M-10000}. \sysname outperforms all the baselines by considerable margins. For poolsize=32 (Tables \ref{tab:bench-3M-32}) and \ref{tab:bench-26M-32}, \sysname outperforms its closest baseline competitor by 0.07 for the MRR metric, and over 10\% for the recall@1 metric. The difference in performance becomes more pronounced when we evaluate the models on the larger pool size of 10,000. For this setup, whose results are presented in Tables \ref{tab:bench-3M-10000} and \ref{tab:bench-26M-10000}, \sysname outperforms its closest competitor by 0.26 for the MRR metric, and over 27\% for the recall@1 metric.

The results demonstrate the merits of our proposed approach, which utilizes both the Transformer architecture and a novel approach for the representation and analysis of the CFG. We can significantly outperform multiple SOTA approaches such as SAFE, which uses an RNN and performs a less rigorous analysis of the CFG, and OrderMatters, which uses Transformer but analyzes each block independently.

Another important aspect of the results is greater relative degradation in the performance of the baselines for larger pool sizes. In the next section we explore this subject further.

\renewcommand\arraystretch{0.9}
\begin{table*}[h!]
\caption{Results of different binary similarity detection methods on \benchname-26M (Poolsize=32)}
\centering
\scalebox{0.85}{
\begin{tabular}{c|ccccccc|ccccccc}
\Xhline{1.5pt}
\hline
             & \multicolumn{7}{c|}{\textbf{MRR}}                             & \multicolumn{7}{c}{\textbf{Recall@1}}                               \\ \hline
\textbf{Models}       & \textbf{O0,O3} & \textbf{O1,O3} & \textbf{O2,O3} & \textbf{O0,Os} & \textbf{O1,Os} & \textbf{O2,Os} & \textbf{Average} & \textbf{O0,O3} & \textbf{O1,O3} & \textbf{O2,O3} & \textbf{O0,Os} & \textbf{O1,Os} & \textbf{O2,Os} & \textbf{Average}  \\ \hline
Gemini      & 0.402          & 0.643          & 0.835          & 0.469          & 0.564         & 0.628         & 0.590           & 0.263          & 0.528         & 0.768          & 0.322          & 0.441          & 0.518          & 0.473          \\
SAFE        & 0.856          & 0.940          & 0.970          & 0.874          & 0.935          & 0.934         & 0.918          & 0.770          & 0.902          & 0.951          & 0.795         & 0.891          & 0.891          & 0.867          \\
Asm2Vec     & 0.439          & 0.847          & 0.958          & 0.490          & 0.788         & 0.849         & 0.729          & 0.314          & 0.789         & 0.940           & 0.362          & 0.716          & 0.800            & 0.654          \\ 
GraphEmb     & 0.583          & 0.681          & 0.741          & 0.610          & 0.637          & 0.639          & 0.649          & 0.465          & 0.586          & 0.667          & 0.499          & 0.541          & 0.543          & 0.550          \\
OrderMatters & 0.572          & 0.820          & 0.932          & 0.630          & 0.692          & 0.771          & 0.729          & 0.417          & 0.740          & 0.903          & 0.481          & 0.692          & 0.677          & 0.652          \\ \hline
\sysname-Zero & 0.632          & 0.871          & 0.973          & 0.687          & 0.890         & 0.891         & 0.824          & 0.539         &  0.838        &  0.961         &  0.602        &   0.854       &    0.853    &       0.775  \\ 
\sysname      & \textbf{0.964} & \textbf{0.983} & \textbf{0.989} & \textbf{0.969} & \textbf{0.980} & \textbf{0.980} & \textbf{0.978} & \textbf{0.941} & \textbf{0.970} & \textbf{0.981} & \textbf{0.949} & \textbf{0.964} & \textbf{0.964} & \textbf{0.962} \\ \hline
\Xhline{1.5pt}
\end{tabular}
}\label{tab:bench-26M-32}
\end{table*}

\begin{table*}[h!]
\renewcommand\arraystretch{0.9}
\caption{Results of different binary similarity detection methods on \benchname-26M (Poolsize=10000)}
\centering
\scalebox{0.85}{
\begin{tabular}{c|ccccccc|ccccccc}
\Xhline{1.5pt}
\hline
             & \multicolumn{7}{c|}{\textbf{MRR}}                             & \multicolumn{7}{c}{\textbf{Recall@1}}                               \\ \hline
\textbf{Models}              & \textbf{O0,O3} & \textbf{O1,O3} & \textbf{O2,O3} & \textbf{O0,Os} & \textbf{O1,Os} & \textbf{O2,Os} & \textbf{Average} & \textbf{O0,O3} & \textbf{O1,O3} & \textbf{O2,O3} & \textbf{O0,Os} & \textbf{O1,Os} & \textbf{O2,Os} & \textbf{Average}  \\ \hline
Gemini      & 0.072          & 0.189          & 0.474          & 0.069          & 0.147          & 0.202         & 0.192          & 0.058          & 0.148          & 0.420          & 0.051         & 0.115          & 0.162          & 0.159          \\
SAFE        & 0.198          & 0.415          & 0.696          & 0.197          & 0.377          & 0.431         & 0.386          & 0.135          & 0.314          & 0.634          & 0.127         & 0.279          & 0.343          & 0.305          \\
Asm2Vec     & 0.118          & 0.443          & 0.703          & 0.107          & 0.369          & 0.480         & 0.370          & 0.099          & 0.376          & 0.638          & 0.086         & 0.307          & 0.413          & 0.320           \\  
GraphEmb     & 0.116          & 0.228          & 0.498          & 0.133          & 0.198          & 0.224          & 0.233          & 0.080          & 0.171          & 0.465          & 0.090          & 0.145          & 0.175          & 0.188          \\
OrderMatters & 0.113          & 0.292          & 0.682          & 0.118          & 0.256          & 0.295          & 0.292          & 0.094          & 0.222          & 0.622          & 0.093          & 0.195          & 0.236          & 0.244          \\ \hline
\sysname-Zero & 0.215          & 0.570          & 0.759          & 0.233          & 0.571          & 0.563         & 0.485          & 0.167          & 0.503          & 0.701          & 0.175         & 0.507          & 0.500          & 0.426           \\
\sysname      & \textbf{0.584} & \textbf{0.734} & \textbf{0.792} & \textbf{0.627} & \textbf{0.709} & \textbf{0.710} & \textbf{0.693} & \textbf{0.499} & \textbf{0.668} & \textbf{0.736} & \textbf{0.550} & \textbf{0.648} & \textbf{0.648} & \textbf{0.625} \\ \hline
\Xhline{1.5pt}
\end{tabular}
}\label{tab:bench-26M-10000}
\end{table*}

\begin{figure*}[h!]
  \centering
  \setlength{\abovecaptionskip}{2mm}
  \includegraphics[width=0.95\linewidth]{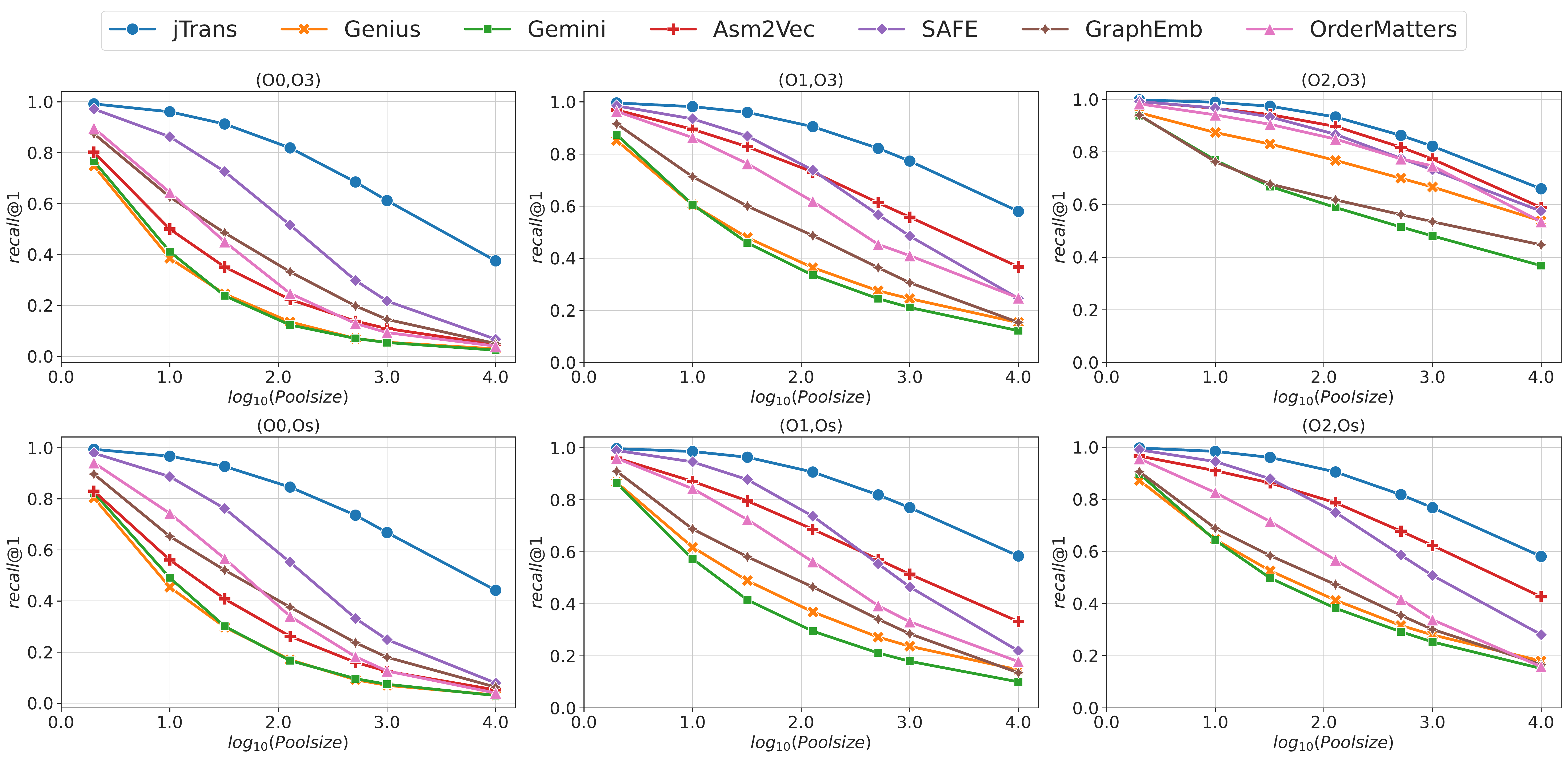}
  \caption{The performance of different binary similarity detection methods on \benchname-3M.}\label{fig:poolsizecompare}
\end{figure*}

\subsection{The Effects of Poolsize on Performance} \label{subsec:RQ2}

The results of the previous section highlight the effect of the poolsize variable on the performance of binary similarity detection algorithms. These results are particularly significant given the fact that previous studies use relatively small poolsizes between 10 and 200, with some studies \cite{xu2017neural,massarelli2019safe}
employing a poolsize of 2. We argue that such setups are problematic, given the fact that for real-world applications such as clone detection and vulnerability search, the poolsize is often larger by orders of magnitude. Therefore, we now present an in-depth analysis of the effects of the poolsize on the performance of SOTA approaches for binary analysis.

The results are presented in Figure \ref{fig:poolsizecompare}. We conducted multiple experiments with a variety of poolsizes---2, 10, 32, 128, 512, 1,000, and 10,000---and plotted the results for various optimization pairs. The results clearly show that all baselines' relative performance is worse than \sysname's as the poolsize increases. Furthermore, our approach does not display sharp drops in its performance (note that the X-axis in Figure \ref{fig:poolsizecompare} is logarithmic), while the baselines' performance generally declines more rapidly once poolsize=100 has been reached. This suggests that our approach is not affected so much by the poolsize as it is by the classification problem becoming more challenging due to a large number of candidates.

Finally, we would like to point out that for a very small poolsizes (e.g., 2), the performance of SOTA baselines such as SAFE and Asm2Vec is almost identical to that of \sysname, with the latter outperforming by approximately 2\%. We deduce that evaluating binary analysis tools on small poolsizes does not provide meaningful indication to their performance in real-world settings.

\subsection{Real-World Vulnerability Search} \label{subsec:RQ3}

Vulnerability detection is considered one of the main applications in computer security. 
We wish to evaluate \sysname's performance on the real-world task of vulnerability search. In this section, we apply \sysname to a known vulnerabilities dataset with the task of searching for vulnerable functions. 

We perform our evaluation on eight CVEs extracted from a known vulnerabilities dataset~\cite{vul}. We produce 10 variants for each function by using different compilers (\texttt{gcc, clang}) and different optimization levels. Our evaluation metric was the recall@10 metric. To simulate real-world settings, we use all of the functions in the project as the search pool. The number of functions for each project varies from 3,038 to 60,159, with the latter being highly challenging. 

\begin{figure}[b]
  \centering
  \setlength{\abovecaptionskip}{0mm}
  \includegraphics[width=0.98\linewidth]{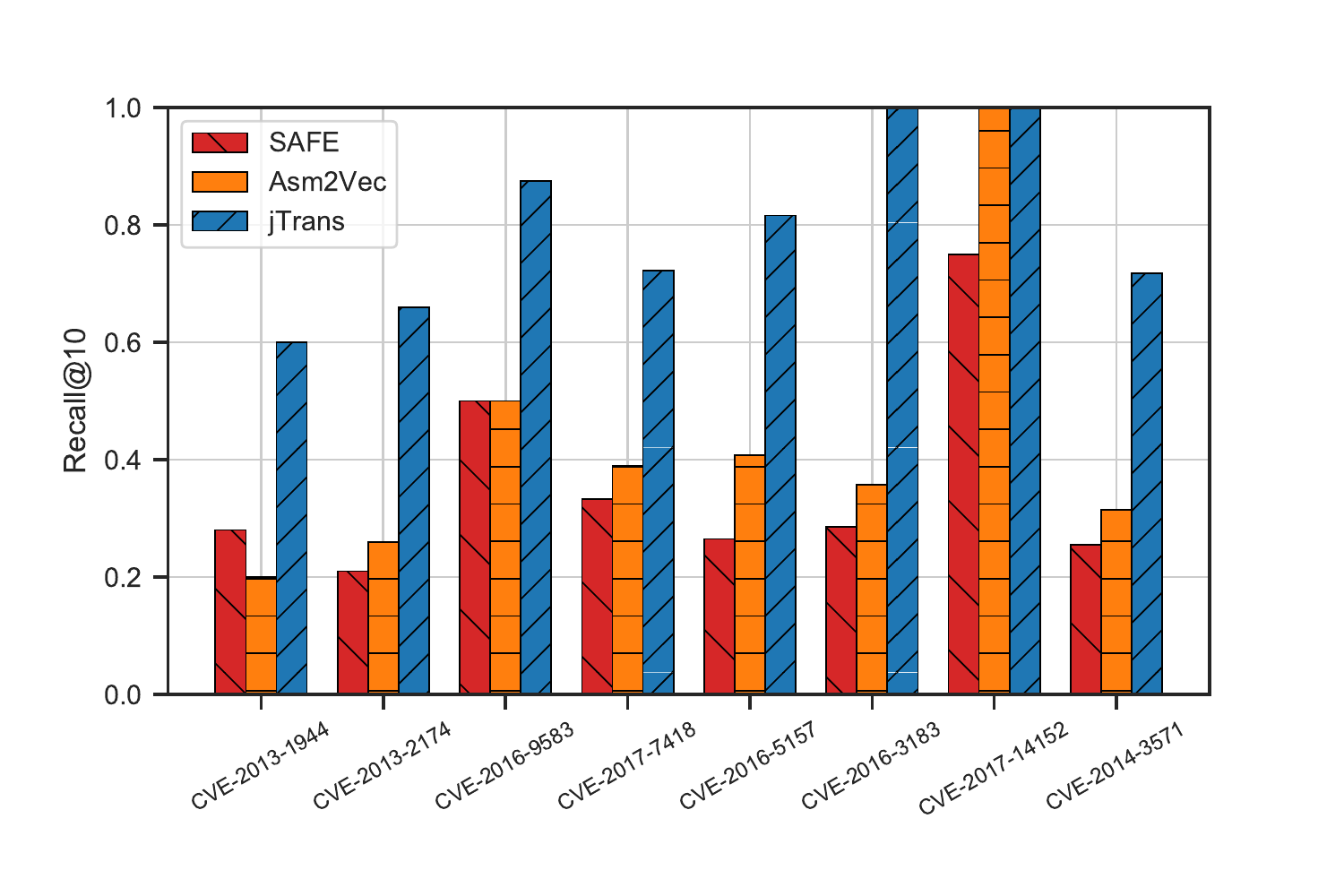}
  \caption{Recall@10 of real-world vulnerability search. }\label{fig:vulsearch}
\end{figure}

Figure \ref{fig:vulsearch} presents the results for the recall@10 metric for each of our queries. We compare our approach to the two leading baselines from Sections \ref{subsec:RQ1} and \ref{subsec:RQ2}---SAFE and Asm2Vec. It is clear that for most of the CVEs, \sysname's performance is significantly higher than that of the two baselines. For example, on \texttt{CVE-2016-3183} from the \texttt{openjpeg} project containing 3,038 functions
, our approach achieved a top-10 recall of $100\%$, meaning that it successfully retrieved all the 10 variants, while Asm2Vec and SAFE achieved recall@10 values of $36.9\%$ and $28.6\%$, respectively. Our results demonstrate that \sysname can be effectively deployed as a vulnerability search tool in real-world scenarios, as a result of its ability to perform well on large pool sizes.

\begin{figure}[b]
  \centering
  \setlength{\abovecaptionskip}{2mm}
  \includegraphics[width=0.9\linewidth]{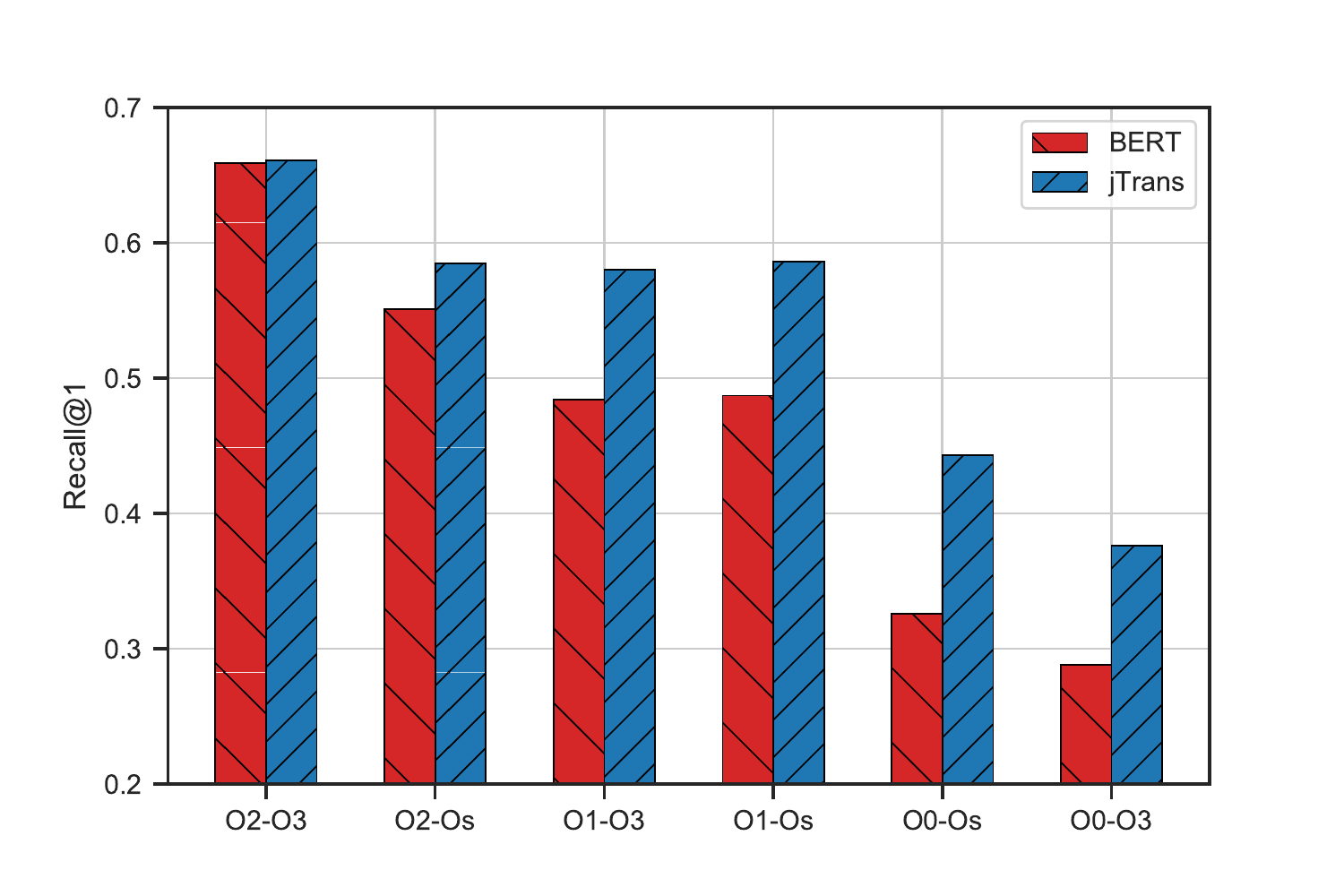}
  \caption{Recall@1 of \sysname and BERT on \benchname-3M when poolsize equals 10,000. }\label{fig:ablation}
\end{figure}

\subsection{The Impact of Our Jump-aware Design}\label{sec:jump_design}
In this section, we test our hypothesis that our jump-aware design significantly contributes to \sysname' ability to analyze the CFG of the binary code. To this end, we train a standard BERT model that does not use our representation of the jump information and compare it to our approach. The hyperparameters used by both models for the pre-training and fine-tuning tasks are identical, with the only distinction being that for BERT we replace the address of each jump token with a fixed token representing an arbitrary address. As a result, similarly to SAFE, the standard  BERT does not receive the control flow information, and can only learn from assembly sequence information.

We evaluate the standard BERT model and \sysname on \benchname-3M using Recall@1, with poolsize=10,000. The results of our evaluation are presented in Figure  \ref{fig:ablation}. 
The results clearly show that the standard BERT performs significantly worse than \sysname, as its performance is lower on each optimization pair. On average, \sysname outperforms BERT by $7.3\%$. These results clearly show that incorporating control flow information into the assembly language sequence modeling is highly beneficial for our model.


\begin{table}[t]
\caption{Accuracy of \sysname in Jump Target Prediction~(JTP) task on \benchname-3M}
\begin{tabular}{ccccc}
\Xhline{1.5pt}
       & \textbf{top-1} & \textbf{top-3} & \textbf{top-5} & \textbf{top-10} \\ \hline
Accuracy & 0.929 & 0.984 & 0.991 & 0.995  \\ 
\Xhline{1.5pt}
\end{tabular}
\label{tab:JTP}
  \vspace{-3mm}
\end{table}

To further explore the efficacy of our jump-aware design, we analyze the ability of our pre-trained model to predict the masked jump addresses in the binary code. We conduct our experiment on \benchname-3M. For each function in the evaluation set, we randomly sample some jump positions with 15\% probability, and replace them with \texttt{<LOC>} in the function. We then analyze the probability of the model correctly predicting the jump target position for each masked jump target. Our results, presented in Table \ref{tab:JTP}, show that \sysname is highly capable at predicting jump positions. Our pre-trained model can predict the target of the jump instruction with top-1 accuracy of 92.9\% and top-10 accuracy of 99.5\%. This accuracy is quite high, particularly for top-1, as there are 512 possible jump positions. These results indicate our pre-trained model was able to successfully capture the contextual instruction information of the binary.

\subsection{Evaluating the Efficacy of Pre-training} \label{subsec:RQ5}
As is the original BERT, pre-training is a critical component of our model. Its main advantage is that it can be performed on unlabeled data, which is much easier to obtain in large quantities. 
To evaluate the effectiveness of the pre-training approach (MLM and JTP), we evaluated a version of our model that \textit{does not perform any fine-tuning}. We follow the same approach as in zero-shot learning~\cite{xian2017zero}, where we use binaries without label information in the pre-training phase. Then, without fine-tuning the pre-trained model, we immediately apply it to the task of binary similarity search. The results of this model, denoted as \sysname-zero, are presented in Tables \ref{tab:bench-3M-32}--\ref{tab:bench-26M-10000}.

The results clearly show the efficacy of our pre-training approach. Even without fine-tuning, \sysname-zero outperforms all the baselines for poolsize=10000: on \benchname-26M, compared to the closest baseline, \sysname-zero improves 0.1 for the MRR metric, and improves 10.6\% for the recall@1 metric. In the poolsize=32 setup, \sysname-zero outperforms all baselines except SAFE, with latter outperforming our approach by 11.4\%. It is important to note, however, that poolsize-32 is far less indicative for real-world scenarios, and in the more challenging poolsize=10000, even our partially-trained approach performed significantly better.



\section{Discussion}

We focus on training \sysname on one architecture (e.g x86) in this paper, but the technique we proposed can be applied to other architectures as well. \sysname provides a novel solution to binary code similarity detection tasks, outperforming state-of-the-art solutions.
It can be applied to many applications, including discovering known vulnerabilities in unknown binaries~\cite{david2014tracelet, pewny2014leveraging, pewny2015cross, eschweiler2016discovre}, 
malware detection~\cite{cesare2013control} and clustering~\cite{hu2009large}, detection of software plagiarism~\cite{saebjornsen2009detecting},  patch analysis~\cite{hu2016cross,xu2017spain}, and software supply chain analysis~\cite{hemel2011finding}.
For instance, due to the rapid deployment of IoT devices, code reusing is very common in IoT development. BCSD solutions like \sysname could help detect whether IoT devices have vulnerabilities revealed in open source libraries.
In the scenario of blockchains, a huge number of blockchains and smart contracts are developed in the past 5 years, based on numerous code cloning and forking. 
However, the security risks of blockchains and smart contracts are severe, and a large portion of them are vulnerable. The code dependency between different blockchains and smart contracts makes this issue even worse. 
We could use \sysname to efficiently detect vulnerabilities in blockchains and smart contracts.

Existing deep learning-based works, as well as \sysname, embed individual binary functions into numerical vectors, and compare similarity between vectors.
As a result, their accuracy drops along with the pool size. As shown in Figure~\ref{fig:poolsizecompare}, the accuracy of most existing solutions drops below 20\% if the pool size is 10,000.
In real world scenarios, the pool size would be much larger.
A model that directly takes two binary functions as input could better 
capture the inter-function relationships and further improve the performance of BCSD, even in a large pool. 
However, training a model to directly compare two functions would have higher overheads. We leave balancing the accuracy and overhead when using \sysname in real world BCSD tasks as a future work.


\section{Conclusion}
    In this work we propose \sysname, the first solution  to embed control-flow information to Transformer-based language models. Our approach utilizes a novel jump-aware architecture design that does not rely on the use of GNNs.
    Theoretical analysis of self attention shows the soundness of our design. Experimental results demonstrate that our method consistently outperforms state-of-the-art approaches by a large margin on BCSD tasks. Through intensive evaluation, we also uncover weaknesses in the evaluation of current SOTA methods. Additionally, we present and release to the community a newly-created dataset named \benchname. Our dataset contains the largest amount of diversified binaries to date, and we believe that it can be used as a high-quality benchmark for future studies in this field. 

\section*{Acknowledgement}
This work was supported in part by National Key R\&D Program of China (2021YFB2701000), National Natural Science Foundation of China under Grant 61972224, Beijing National Research Center for Information Science and Technology under Grant BNR2022RC01006, and Ant Group through CCF-Ant Innovative Research Program No. RF20210021. We would like to thank Jingwei Yi and Bolun Zhang for their great comments and help on experiments.

\appendix
\section*{Appendix}
 \begin{proof}\label{proof_jump_embedding}
We denote the embedding of the $m$-th layer as $E_m$, we first project the $m$-th embedding to $Q_m$, $K_m$ and $V_m$, respectively. 
Then we used the scaled dot-product attention to get the attention matrix $\text{Attention}$. 
   For notation brevity, we use $E,Q,K,V$ to replace $E_m,Q_m,K_m,V_m$, use $q,k,v,d$ to replace $W^Q_m, W^K_m, W^V_m, d_k$ in the derivation.
   
   From the embedding projection and attention calculation, we have:
   
     \begin{equation}
   \begin{aligned}
   Q=E\times q, \quad
   K=E\times k, \quad
   V=E\times v\\
   \text{Attention}(Q,K, V) = \text{Softmax}(\frac{Q K^T}{\sqrt{d}}) \cdot V \\
   \end{aligned}
    \end{equation}
    
    We denote the attention weight matrix before softmax is $A$:
    
     \begin{equation}
    A=\frac{Q\times K^{T}}{\sqrt{d}}=\frac{Eq\times k^{T}E^{T}}{\sqrt{d}}\\
    \end{equation}
    
   Let matrix $J=\frac{q\times k^{T}}{\sqrt{d}}$, then we have: 
   
     \begin{equation}
    A=E J  E^{T}\\
    \end{equation}
     \begin{equation}
    \forall i,j, A_{i,j}=E_i J (E_j)^T=tr(E_i J (E_j)^T)=tr((E_j)^{T} E_i J) \\
    \end{equation}
    
   We then apply SVD decomposition to matrix $J$, and get
   $J=U\times S \times V^{T}$, where, $U$ and $V$ are orthogonal, $S=diag(s_1,s_2,...,s_N), s_i\ge 0$.
   \begin{align}
   A_{i,j}&=tr((E_j)^{T} E_i U S V^{T})\\
   &=tr((E_j)^{T} E_i U S)\\
   &=tr(S (E_j)^{T} E_i U)\\
   &=tr(S (E_j)^{T} E_i)\\
   &=tr(diag(s_1,s_2,...,s_N) (E_j)^{T} E_i)\\
   &=\sum_{n=1}^N s_n  E_{in} E_{jn}
   \end{align} 
   
   Similarly, we could get
   
    \begin{equation}
   \forall l, A_{il}=\sum_{n=1}^N s_n  E_{in} E_{ln}
   \end{equation}
   
   Therefore we have 
    \begin{equation}
    A_{ij}-A_{il}=\sum_{n=1}^N s_n E_{in} (E_{jn}-E_{ln})
    \end{equation}
   
   Denote $T$ as token embedding of current layer, $P$ is position embedding of current layer. Assume $E=P+T$. When position $i$ is jump connected with position $j$. $T_i=P_j$, other embeddings follow a normal distribution
   \begin{align}
       &T_i \sim N(0,\sigma^{2}I), P_i \sim N(0,\sigma^{2}I), \\
       &T_j \sim N(0,\sigma^{2}I), \\
       &T_l \sim N(0,\sigma^{2}I), P_l \sim N(0,\sigma^{2}I)
   \end{align} 
   
   We have 
    \begin{equation}
    A_{ij}-A_{il}=\sum_{n=1}^N s_n (P_{in}+T_{in}) (T_{jn}+P_{jn}-T_{ln}-P_{ln})
    \end{equation}
    
   Taking the expectation of this formula,
   
   \begin{flalign}
   \mathbb{E}(A_{ij}-A_{il})&=\sum_{n=1}^N s_n \mathbb{E}((P_{in}+T_{in}) (T_{jn}+P_{jn}-T_{ln}-P_{ln}))\\
                       &=\sum_{n=1}^N s_n \mathbb{E}(T_{in} P_{jn})\\
                       &=\sum_{n=1}^N s_n \mathbb{E}(T_{in}^2)\\
                       &=\sigma^2 \sum_{n=1}^N s_n
    \end{flalign}
   
   Therefore we have 
    \begin{equation}
    \mathbb{E}(A_{ij}-A_{il})=\sigma^2 \sum_{n=1}^N s_n >0
    \end{equation}
   
 \end{proof}

\bibliographystyle{ACM-Reference-Format}
\bibliography{sample-base}

\end{document}